\documentclass[aip,apl,amsmath,amssymb,reprint,superscriptaddress]{revtex4-1}
\usepackage{graphicx}
\usepackage{dcolumn}
\usepackage{bm}
\usepackage{color}
\begin{document}
\title{Nitrogen-vacancy centers created by N$^+$ ion implantation through screening SiO$_2$ layers on diamond}
\author{Kazuki Ito}
\affiliation{School of Fundamental Science and Technology, Keio University, 3-14-1 Hiyoshi, Kohoku-ku, Yokohama 223-8522, Japan}
\author{Hiroshi Saito}
\affiliation{School of Fundamental Science and Technology, Keio University, 3-14-1 Hiyoshi, Kohoku-ku, Yokohama 223-8522, Japan}
\author{Hideyuki Watanabe}
\affiliation{Correlated Electronics Group, Electronics and Photonics Research Institute, National Institute of Advanced Industrial Science and Technology (AIST),
Tsukuba Central 5, 1-1-1 Higashi, Tsukuba, Ibaraki 305-8565, Japan}
\author{Tokuyuki Teraji}
\affiliation{\mbox{National Institute for Materials Science, 1-1 Namiki, Tsukuba, Ibaraki 305-0044, Japan}}
\author{Kohei M. Itoh}
\email{kitoh@appi.keio.ac.jp}
\affiliation{School of Fundamental Science and Technology, Keio University, 3-14-1 Hiyoshi, Kohoku-ku, Yokohama 223-8522, Japan}
\affiliation{\mbox{Spintronics Research Center, Keio University, 3-14-1 Hiyoshi, Kohoku-ku, Yokohama 223-8522, Japan}}
\author{Eisuke Abe}
\email{e-abe@keio.jp}
\affiliation{\mbox{Spintronics Research Center, Keio University, 3-14-1 Hiyoshi, Kohoku-ku, Yokohama 223-8522, Japan}}
\date{\today}

\begin{abstract}
We report on an ion implantation technique utilizing a screening mask made of SiO$_2$ to control both the depth profile and the dose.
By appropriately selecting the thickness of the screening layer, this method fully suppresses the ion channeling,
brings the location of the highest NV density to the surface, and effectively reduces the dose by more than three orders of magnitude.
With a standard ion implantation system operating at the energy of 10~keV and the dose of 10$^{11}$~cm$^2$ and without an additional etching process,
we create single NV centers close to the surface with coherence times of a few tens of $\mu$s.
\end{abstract}
\maketitle

Impurity doping of semiconductors is a fabrication process indispensable for the modern electronic devices, and continues to be so for quantum devices
such as silicon-based single-donor spin qubits~\cite{K98,PTD+12,MDL+14}, in which the positions of the individual donors must be controlled precisely.~\cite{DYA+15}
Being optically addressable and coherently controllable by microwaves,
the single electronic spins associated with the negatively charged nitrogen-vacancy (NV$^-$) centers in diamond are playing important roles in emergent quantum technology,
{\it e.g.,} as a matter qubit interfacing with a flying qubit~\cite{TCT+10,BHP+13,PHB+14,HBD+15} and as a nanoscale magnetic sensor.~\cite{MSH+08,BCK+08,RTH+14}
Both applications demand that the NV centers be located close to the diamond surface.~\cite{WJGM13}
For quantum network, shallow NV spins can be efficiently coupled to photons in a nanophotonic cavity.~\cite{SMZ+16}
For magnetometry, the proximity of the NV sensor to a magnetic specimen is crucial, because their dipolar coupling strength decays as the inverse cube of the separation.

So far, shallow NV centers ($<$ 5~nm from the surface) have been created primarily by
(i) nitrogen-doping during CVD growth~\cite{OHB+12,ORW+13,OHB+14,KSB+16}
and (ii) N$^+$ ion implantation.~\cite{MBD+05,RRT+06,TWF+10,SDR+11,PRB+11,SBAP14,PNJ+10,OPC+12,OSP+13,AHA+14}
The CVD approach allows the accurate control of the impurity distribution in the depth direction, whereas the doping is random in the lateral dimensions.
Ironically, high-quality CVD diamond films tend to lack vacancies to pair up with nitrogen atoms;
an additional process to introduce vacancies, such as electron irradiation~\cite{OHB+12}, C$^+$ ion implantation~\cite{OHB+14}, or He$^+$ ion implantation~\cite{KSB+16}, is often required,
although the creation of shallow NV centers in as-grown films has also been reported.~\cite{ORW+13}

Ion implantation introduces both nitrogen atoms and vacancies into diamond.
The lateral distributions are controllable by the use of focused ion beam~\cite{MBD+05} or an array of small apertures.~\cite{TWF+10,SDR+11,PRB+11,SBAP14} 
The main concern is that the depth profile intrinsically has broadening approximated by a Gaussian distribution.
Importantly, the ions can penetrate deep inside of the crystal lattice due to the ion channeling effect.~\cite{SL12}
The prevailing approach is to keep the implantation energy low ($<$ 5~keV).~\cite{PNJ+10,OPC+12,OSP+13,AHA+14}
It is also preferred to set the implantation dose (fluence) low ($\sim$10$^{8}$~cm$^{-2}$), so that single NV centers can be resolved optically.
On the other hand, standard, multi-purpose ion implantation systems operate at 10~keV or higher with the dose of 10$^{11}$~cm$^{-3}$ or higher,
making this approach not widely available.
Even with high-energy ion implantation, it is possible to plasma-etch the implanted diamond to bring the NV centers closer to the surface.
But special cares must be taken to prevent the etching itself from damaging the surface.~\cite{CGO+15,OMW+15}

In this paper, we report on an ion implantation technique utilizing a screening mask made of SiO$_2$ to control both depth profile and dose.
With a standard ion implantation system operating at the energy of 10~keV and the dose of 10$^{11}$~cm$^2$ and without additional surface-etching, we create single NV centers close to the surface.
The use of a screening mask is quite common in silicon industry~\cite{SL12}, and yet has not been explored thoroughly in the context of NV centers.
Previously, thin ($<$ 10~nm) or thick ($\sim$100~nm) screening layers were employed to mitigate the ion channeling or fully suppress the ion transmission into diamond~\cite{TWF+10,SDR+11,PRB+11,SBAP14},
but not to control the entire depth profile.
We show here that the appropriate selection of screening layer thickness has a qualitatively different consequence that is preferable to create shallow NV centers.

\begin{figure}
\begin{center}
\includegraphics{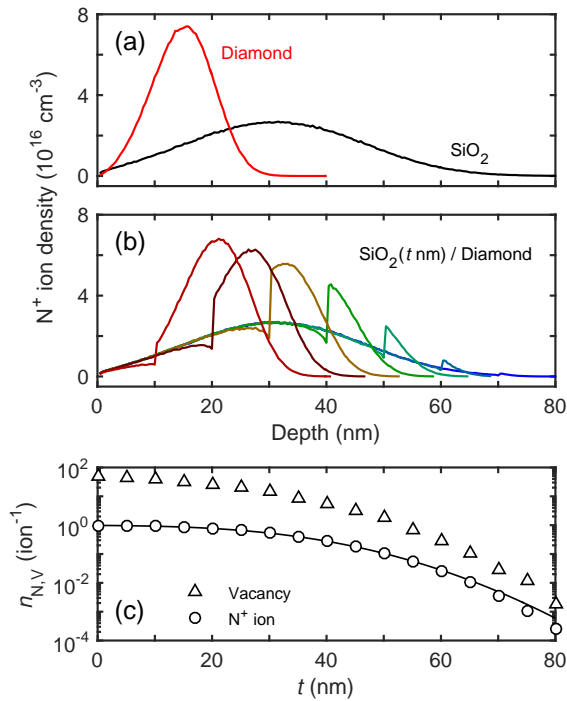}
\caption{Monte Carlo simulations (SRIM) of (a) N$^+$ ion implantations into diamond and SiO$_2$, and (b) SiO$_2$ ($t$~nm)/diamond heterostructures with $t$ = 10, 20, 30, 40, 50, 60, and 70~nm.
The interfaces are discernible as vertical lines.
(c) $n_{\mathrm{N}}$ ($\bigcirc$) and $n_{\mathrm{V}}$ ($\triangle$) as functions of $t$.
The solid line is a calculation using Eq.~(\ref{n_in_diamond})
\label{fig1}}
\end{center}
\end{figure}

To illustrate our approach, we carry out Monte Carlo simulations of ion implantation using a software package SRIM.~\cite{SRIM,Z04}
Figure~\ref{fig1}(a) shows the simulation results for diamond and SiO$_2$ with the N$^+$ energy of 10~keV and the dose of 10$^{11}$~cm$^{-2}$.
The incident angle is normal to the surface.
Fitting the curves by Gaussian distribution functions,
we obtain the {\it projected range} $R$ (the mean travel distance of ions along the axis of the incidence) and the {\it projected straggle} $\sigma$ (standard deviation) as
$R_{\mathrm{d}}$ = 15.0~nm and $\sigma_{\mathrm{d}}$ = 5.4~nm for diamond,
and
$R_{\mathrm{m}}$ = 31.1~nm and $\sigma_{\mathrm{m}}$ = 15.1~nm for SiO$_2$, respectively.
We note that SRIM does not take into account the ion channeling and therefore the actual depth profile for diamond can have a longer tail toward the interior of the material.
On the other hand, SiO$_2$ is an amorphous material expected to be free of the ion channeling.

We then simulate the case of SiO$_2$ ($t$~nm)/diamond heterostructures.
Figure~\ref{fig1}(b) shows the results for various $t$ ranging between 10 and 70~nm with 10~nm steps.
The depth profiles now appear as combinations of the profiles of the two materials;
down to the SiO$_2$/diamond interfaces at $t$~nm, the profiles trace that of SiO$_2$ (except for the small reductions near the interfaces),
and after entering into diamond the profiles mimic that of diamond with the near-surface parts truncated.
Qualitatively, these profiles are analogous to the case in which N$^+$-ion-implanted diamond (without SiO$_2$) is surface-etched by
$d_{\mathrm{e}}$ = $(\sigma_{\mathrm{d}}/\sigma_{\mathrm{m}}) (t - R_{\mathrm{m}}) + R_{\mathrm{d}}$~nm,
where $d_{\mathrm{e}}$ is determined so that the integration of the profile for SiO$_2$ larger than $t$ and that for diamond larger than $d_{\mathrm{e}}$ may be equal.
Notably, for $t \geq$ 40~nm ($> R_{\mathrm{m}}$), only the tail parts appear in the profiles inside of diamond, thereby the locations of the highest ion densities are {\it at the surface}.
It is also clear that the distributions in the depth direction become much narrower than $\sigma_{\mathrm{d}}$.
In addition, the simulated profiles should be more reliable for larger $t$, owing to the better suppression of ion channeling by SiO$_2$ layers.

In Fig.~\ref{fig1}(c), we plot the fraction of N$^+$ ions transmitted into diamond $n_{\mathrm{N}}$ and the average number of vacancies in diamond created by a single ion $n_{\mathrm{V}}$.
$n_{\mathrm{N}} (t)$ can also be estimated as~\cite{SL12}
\begin{equation}{
n_{\mathrm{N}} (t) = \frac{1}{2} \mathrm{erfc} \left( \frac{ t - R_{\mathrm{m}} }{ \sqrt{2} \sigma_{\mathrm{m}} } \right),
\label{n_in_diamond}}
\end{equation}
reproducing well the simulation results.
Figure~\ref{fig1}(c) shows that we are able to reduce the {\it effective} dose by three orders of magnitude, at the same time reducing the number of vacancies by nearly four orders of magnitude.
The excess vacancies not diffused away or annihilated by thermal annealing after the ion implantation are the sources to degrade the crystalline quality;
creating less vacancies at the stage of ion implantation can be beneficial.
Assuming a moderate N-to-NV conversion efficiency (yield) of 1~\%, the final NV density of $\sim$10$^{11}$~cm$^{-3}$ (or $\sim$2 $\times$ 10$^{7}$~cm$^{-2}$),
favorable to the single NV detection, is attainable.

The benefits of using a screening mask with appropriately-chosen thickness are summarized as follows;
(i) the ion channeling is fully suppressed,
(ii) the location of the highest NV density can be brought to the surface, and
(iii) the effective dose can be reduced by more than three orders of magnitude.

We now test our approach experimentally.
The sample was a natural abundant, (100)-oriented HPHT IIa diamond (Sumitomo), on which multiple SiO$_2$ layers with different thickness were deposited by electron beam evaporation.
During the SiO$_2$ deposition and the N$^+$ ion implantation, the sample was kept covered with a metal plate equipped with apertures with the diameters of 400~$\mu$m.
In each deposition run, only one aperture was open, and the rest of the apertures were blocked.
On the sample stage of the evaporator, a silicon substrate was also set in the vicinity of the sample.
After each run, we determined the thickness $t$ of deposited SiO$_2$ by ellipsometry to the silicon substrate (the thickness of native oxide is taken into account).
The sample was then implanted with $^{15}$N$^+$ ions at 10~keV and the dose $D_{\mathrm{N}}$ = 10$^{11}$~cm$^{-2}$ (Ion Technology Center), with all the apertures open.
The $^{15}$N isotopes ($I$ = $\frac{1}{2}$) were used to discriminate from the $^{14}$N isotopes in the bulk ($I$ = 1, 99.6~\%).
After the ion implantation, the SiO$_2$ layers were removed by hydrofluoric acid.
The sample was subsequently annealed at 800~$^{\circ}$C for 2~h in vacuum in order to let vacancies diffuse to form NV centers,
and at 450~$^{\circ}$C for 9~h in oxygen atmosphere in order to convert neutral NV (NV$^0$) centers into negatively charged ones (NV$^-$).~\cite{FSB+10}
The sample was cleaned before and after each annealing process.
In the following, the areas with $t$ = 0, 16, 25, 36, 46, 53, 64, and 72~nm are measured in detail.

\begin{figure*}
\begin{center}
\includegraphics{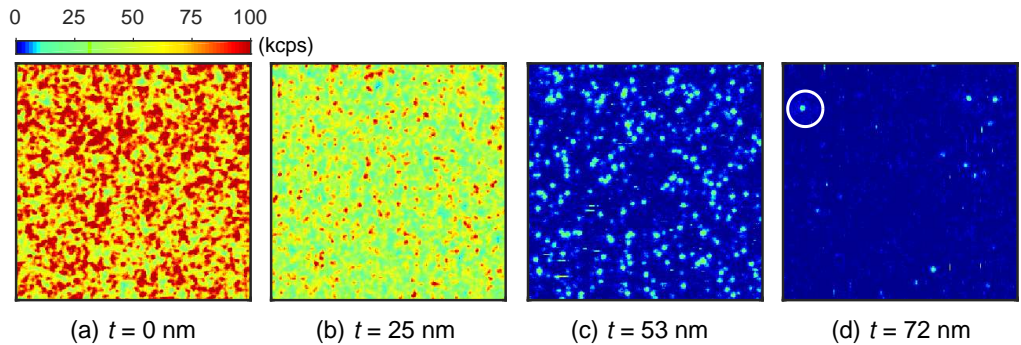}
\caption{Fluorescence images (20~$\times$~20~$\mu$m$^2$) of areas N$^+$-ion-implanted through SiO$_2$ layers with $t$ = 0~nm (a), 25~nm (b), 53~nm (c), and 72~nm (d).
The white circle indicates the spot from which the data in Fig.~\ref{fig4}(a--c) were taken.
\label{fig2}}
\end{center}
\end{figure*}

The properties of the created NV centers are examined by fluorescence imaging and optically-detected magnetic resonance (ODMR),
using a home-built confocal microscope combined with microwave circuitry.
The photons emitted from the NV centers (600--800~nm) are collected by a single-photon counting module,
and our setup typically gives $\sim$25~kcps (kilo-counts per second) photons from a single NV center located close to a diamond surface,
which is calibrated using a different diamond sample containing near-surface NV centers (similar to the one reported in Ref.~\onlinecite{ORW+13}).
Figure~\ref{fig2} shows representative fluorescence images taken at different areas of the sample surface.
As increasing $t$, the photon counts from the surface progressively decrease.
For the areas with $t \leq$ 36~nm, we observe ensemble NV centers.
In the $t$ = 46~nm area, both single and ensemble NV centers are observed.
And for the areas with $t \geq$ 53~nm, large portions of the images are dark, with discrete spots emitting a few tens kcps photons as expected for single NV centers.
We note that, in claiming NV centers (single or ensemble) here, they are confirmed to originate from $^{15}$NV centers [see Fig.~\ref{fig4}].
We also note that from photoluminescence (PL) spectroscopy we have not detected signals unrelated to NV centers, for instance silicon-vacancy (SiV) centers.
It is possible that our method introduces Si atoms into diamond due to a knock-on effect and subsequently produces SiV centers.
However, as far as we checked, such SiV centers have not been found, and we conclude that the knock-on of Si atoms during the N$^+$ ion implantation is negligible here.

\begin{figure}[b]
\begin{center}
\includegraphics{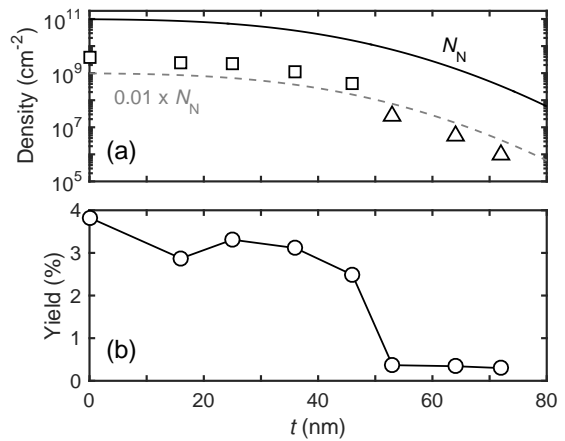}
\caption{(a) The experimental NV density $N_{\mathrm{NV}}$ and the calculated N density $N_{\mathrm{N}}$, 0.01 $\times$ $N_{\mathrm{N}}$ as functions of $t$.
The square (triangle) points $\square$ ($\triangle$) are estimated from ensemble (single) NV centers. 
(b) Yield ($N_{\mathrm{NV}}/N_{\mathrm{N}}$) as a function of $t$.
\label{fig3}}
\end{center}
\end{figure}

We estimate the NV density $N_{\mathrm{NV}}$ in two ways.
For the areas with $t \leq$ 46~nm, where ensemble NV centers are observed,
we calculate the average photon counts per laser spot size and divide them by the calibrated photon counts from a single NV center (25~kpcs).
As the depth resolution of our confocal microscope is at most a micron, much larger than the expected depth profile,
we calculate the average NV number in the unit area (in cm$^{-2}$ unit).
For the areas with $t \geq$ 53~nm, we count the number of bright spots in the observed region.
To correctly identify single NV$^-$ centers and exclude the signals from NV$^0$ as well as other spurious emitters such as surface dusts,
the threshold values for photon counts and spot size are defined.
The estimated $N_{\mathrm{NV}}$ is shown in Fig.~\ref{fig3}(a).
For comparison, the N density $N_{\mathrm{N}}$ calculated as $D_{\mathrm{N}} \times n_{\mathrm{N}}$ is drawn. 
Also drawn is 0.01 $\times$ $N_{\mathrm{N}}$, which roughly follows the $N_{\mathrm{NV}}$ data points, indicating the yield of about 1\%.
Figure~\ref{fig3}(b) plots the yield calculated as $N_{\mathrm{NV}}/N_{\mathrm{N}}$.
It is observed that the yields for single NV centers are $\sim$0.2~\%, less efficient than those for ensemble NV centers ($\sim$3\%).
The better values in the ensemble case may partly be attributed to the inclusion of the phonon side band of NV$^0$ in the photon counts.
Extending up to 700~nm, the phonon side band of NV$^0$ partially overlaps with the NV$^-$ spectrum.
However, from PL spectroscopy, we estimate the ratio NV$^-$/(NV$^-$ + NV$^0$) to be around 80\%;
this is not sufficient to explain the difference in the yields.
It is possible that the reduced number of vacancies for larger $t$ [see Fig~\ref{fig2}(c)] resulted in the reduced probabilities of paring up a nitrogen atom and a vacancy.
A further work is necessary to understand the $t$-dependence of the yield.

Next, we examine the spin properties of single NV centers found in the $t \geq$ 46~nm areas.
An example of the ODMR spectrum taken at the $t$ = 72~nm area is shown in Fig.~\ref{fig4}(a).
\begin{figure}
\begin{center}
\includegraphics{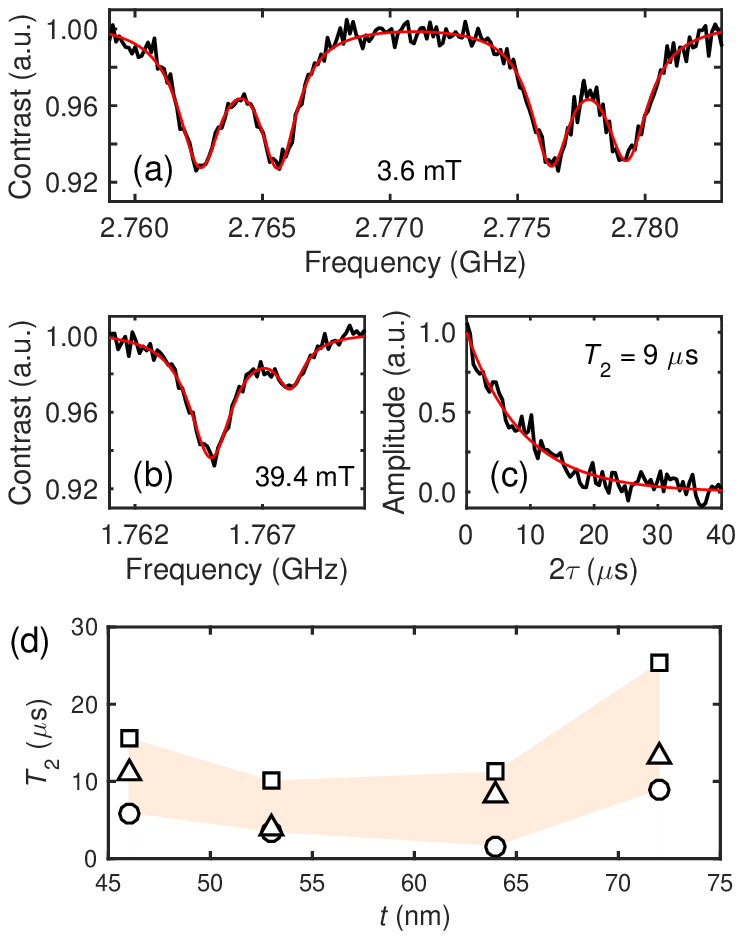}
\caption{(a) ODMR spectrum of a single NV center found at the $t$ = 72~nm area [the white circle in Fig.~\ref{fig2}(d)], exhibiting the hyperfine couplings with $^{13}$C and $^{15}$N nuclei.
(b) ODMR spectrum of the same NV center as in (a) at higher magnetic fields. 
(c) Hahn echo decay curve taken at the condition of (b).
(d) $T_2$ measured for different single NV centers in various areas.
\label{fig4}}
\end{center}
\end{figure}
The resonance dip corresponding to the $m_S$ = 0 $\leftrightarrow$ $-$1 transition is split by 14~MHz due to the hyperfine coupling with a proximal $^{13}$C nucleus ($I$ = $\frac{1}{2}$, 1\%),
attributed to the third shell.~\cite{GFK08,MNR+09}
The dips are further split by 3~MHz due to the hyperfine coupling with $^{15}$N nucleus of its own.
Note that the $m_S$ = 0 $\leftrightarrow$ 1 transitions are at 200~MHz higher frequencies and are not shown.
In addition to giving  $\sim$25~kcps photons, the coupling to a specific $^{13}$C nucleus in the lattice evidences that the observed spectrum arises from a single NV center.
The $^{15}$N hyperfine interaction unambiguously tells that the NV center is formed from the implanted nitrogen.

In measuring the coherence time $T_2$ of the NV spins by a Hahn-echo sequence~\cite{H50}, two factors contributing the echo signal must be considered.
One is the electron spin echo envelope modulation (ESEEM) effect due to the $^{15}$N nucleus of the NV center~\cite{OHB+12}, and the other is the echo revivals due to $^{13}$C nuclei in the bulk.~\cite{CDT+06}
To mitigate them, we work at higher magnetic fields, as shown in Fig.~\ref{fig4}(b).
Aligned along the NV axis, the static magnetic field of a few tens of mT serves to polarize the $^{15}$N nucleus and suppress the $^{15}$N ESEEM.
In addition, the revival period of the $^{13}$C nuclear oscillation in this case is a few $\mu$s, thus cannot be overlooked even when $T_2$ is short.
Figure~\ref{fig4}(c) shows the echo decay curve obtained at the condition of (b), giving $T_2$ = 9~$\mu$s.
We repeat the $T_2$ measurements for various single $^{15}$NV centers, and the results are summarized in Fig.~\ref{fig4}(d).
The values of $T_2$ vary from one NV center to another, but they fall on about 10~$\mu$s.
The longest $T_2$ we have obtained so far is 23~$\mu$s from the $t$ = 72~nm area.

To summarize, we have proposed and demonstrated an ion implantation technique utilizing a screening mask made of SiO$_2$ to control both depth profile and dose.
Even with the implantation energy of 10~keV and the dose of 10$^{11}$~cm$^{2}$, which are in general too high to create shallow single NV centers,
we are able to create single NV centers with $T_2$ of a few tens of $\mu$s.
In addition to being simple and less demanding in the sense that neither a custom-built low-energy low-dose ion implantation system nor a careful plasma-etching process is required,
this method is easily combined with the aperture array method to provide additional control in the lateral distribution.
We also point out that a material for the screening mask is flexible and not limited to SiO$_2$;
One can use other materials as long as their thicknesses are precisely controllable and they are preferably amorphous.
AlO$_2$ or organic resists may be used, for instance.
Future works include the determination of the depth profile of the NV centers by using, for instance, proton nuclear magnetic resonance (NMR) of external, surface nuclear spins.~\cite{PDC+16}
The present diamond substrate contains $^{13}$C nuclei, the NMR signal of which can mimic that of protons.~\cite{LBR+16}
The use of isotopically pure $^{12}$C diamond as a starting material will avoid this problem and can also improve the coherence properties of the NV spins.
Relatively short $T_2$ observed here is indicative of the near-surface nature of the NV centers created,
but it should also be noted that the nitrogen and vacancy distributions can be quite different from the standard ion implantation.
To understand the formation dynamics of the NV centers both qualitatively and quantitatively, a combination of molecular dynamics and kinetic Monte Carlo simulations will be important.~\cite{dOAW+17}

We thank H. Sumiya for supplying the diamond substrate.
HW acknowledges support from JSPS Grant-in-Aid for Scientific Research (KAKENHI) (A) No.~26249108 and JST Development of Systems and Technologies for Advanced Measurement and Analysis (SENTAN).
TT acknowledges support from KAKENHI (B) No. 15H03980.
KMI acknowledges support from KAKENHI (S) No.~26220602, JSPS Core-to-Core Program, and Spintronics Research Network of Japan (Spin-RNJ).
\bibliography{ionimp}

\begin{thebibliography}{40}%
\makeatletter
\providecommand \@ifxundefined [1]{%
 \@ifx{#1\undefined}
}%
\providecommand \@ifnum [1]{%
 \ifnum #1\expandafter \@firstoftwo
 \else \expandafter \@secondoftwo
 \fi
}%
\providecommand \@ifx [1]{%
 \ifx #1\expandafter \@firstoftwo
 \else \expandafter \@secondoftwo
 \fi
}%
\providecommand \natexlab [1]{#1}%
\providecommand \enquote  [1]{``#1''}%
\providecommand \bibnamefont  [1]{#1}%
\providecommand \bibfnamefont [1]{#1}%
\providecommand \citenamefont [1]{#1}%
\providecommand \href@noop [0]{\@secondoftwo}%
\providecommand \href [0]{\begingroup \@sanitize@url \@href}%
\providecommand \@href[1]{\@@startlink{#1}\@@href}%
\providecommand \@@href[1]{\endgroup#1\@@endlink}%
\providecommand \@sanitize@url [0]{\catcode `\\12\catcode `\$12\catcode
  `\&12\catcode `\#12\catcode `\^12\catcode `\_12\catcode `\%12\relax}%
\providecommand \@@startlink[1]{}%
\providecommand \@@endlink[0]{}%
\providecommand \url  [0]{\begingroup\@sanitize@url \@url }%
\providecommand \@url [1]{\endgroup\@href {#1}{\urlprefix }}%
\providecommand \urlprefix  [0]{URL }%
\providecommand \Eprint [0]{\href }%
\providecommand \doibase [0]{http://dx.doi.org/}%
\providecommand \selectlanguage [0]{\@gobble}%
\providecommand \bibinfo  [0]{\@secondoftwo}%
\providecommand \bibfield  [0]{\@secondoftwo}%
\providecommand \translation [1]{[#1]}%
\providecommand \BibitemOpen [0]{}%
\providecommand \bibitemStop [0]{}%
\providecommand \bibitemNoStop [0]{.\EOS\space}%
\providecommand \EOS [0]{\spacefactor3000\relax}%
\providecommand \BibitemShut  [1]{\csname bibitem#1\endcsname}%
\let\auto@bib@innerbib\@empty
\bibitem [{\citenamefont {Kane}(1998)}]{K98}%
  \BibitemOpen
  \bibfield  {author} {\bibinfo {author} {\bibfnamefont {B.~E.}\ \bibnamefont
  {Kane}},\ }\href@noop {} {\bibfield  {journal} {\bibinfo  {journal} {Nature}\
  }\textbf {\bibinfo {volume} {393}},\ \bibinfo {pages} {133} (\bibinfo {year}
  {1998})}\BibitemShut {NoStop}%
\bibitem [{\citenamefont {Pla}\ \emph {et~al.}(2012)\citenamefont {Pla},
  \citenamefont {Tan}, \citenamefont {Dehollain}, \citenamefont {Lim},
  \citenamefont {Morton}, \citenamefont {Jamieson}, \citenamefont {Dzurak},\
  and\ \citenamefont {Morello}}]{PTD+12}%
  \BibitemOpen
  \bibfield  {author} {\bibinfo {author} {\bibfnamefont {J.~J.}\ \bibnamefont
  {Pla}}, \bibinfo {author} {\bibfnamefont {K.~Y.}\ \bibnamefont {Tan}},
  \bibinfo {author} {\bibfnamefont {J.~P.}\ \bibnamefont {Dehollain}}, \bibinfo
  {author} {\bibfnamefont {W.~H.}\ \bibnamefont {Lim}}, \bibinfo {author}
  {\bibfnamefont {J.~J.~L.}\ \bibnamefont {Morton}}, \bibinfo {author}
  {\bibfnamefont {D.~N.}\ \bibnamefont {Jamieson}}, \bibinfo {author}
  {\bibfnamefont {A.~S.}\ \bibnamefont {Dzurak}}, \ and\ \bibinfo {author}
  {\bibfnamefont {A.}~\bibnamefont {Morello}},\ }\href@noop {} {\bibfield
  {journal} {\bibinfo  {journal} {Nature}\ }\textbf {\bibinfo {volume} {489}},\
  \bibinfo {pages} {541} (\bibinfo {year} {2012})}\BibitemShut {NoStop}%
\bibitem [{\citenamefont {Muhonen}\ \emph {et~al.}(2014)\citenamefont
  {Muhonen}, \citenamefont {Dehollain}, \citenamefont {Laucht}, \citenamefont
  {Hudson}, \citenamefont {Kalra}, \citenamefont {Sekiguchi}, \citenamefont
  {Itoh}, \citenamefont {Jamieson}, \citenamefont {McCallum}, \citenamefont
  {Dzurak},\ and\ \citenamefont {Morello}}]{MDL+14}%
  \BibitemOpen
  \bibfield  {author} {\bibinfo {author} {\bibfnamefont {J.~T.}\ \bibnamefont
  {Muhonen}}, \bibinfo {author} {\bibfnamefont {J.~P.}\ \bibnamefont
  {Dehollain}}, \bibinfo {author} {\bibfnamefont {A.}~\bibnamefont {Laucht}},
  \bibinfo {author} {\bibfnamefont {F.~E.}\ \bibnamefont {Hudson}}, \bibinfo
  {author} {\bibfnamefont {R.}~\bibnamefont {Kalra}}, \bibinfo {author}
  {\bibfnamefont {T.}~\bibnamefont {Sekiguchi}}, \bibinfo {author}
  {\bibfnamefont {K.~M.}\ \bibnamefont {Itoh}}, \bibinfo {author}
  {\bibfnamefont {D.~N.}\ \bibnamefont {Jamieson}}, \bibinfo {author}
  {\bibfnamefont {J.~C.}\ \bibnamefont {McCallum}}, \bibinfo {author}
  {\bibfnamefont {A.~S.}\ \bibnamefont {Dzurak}}, \ and\ \bibinfo {author}
  {\bibfnamefont {A.}~\bibnamefont {Morello}},\ }\href@noop {} {\bibfield
  {journal} {\bibinfo  {journal} {Nat.\ Nanotechnol.}\ }\textbf {\bibinfo
  {volume} {9}},\ \bibinfo {pages} {986} (\bibinfo {year} {2014})}\BibitemShut
  {NoStop}%
\bibitem [{\citenamefont {van Donkelaar}\ \emph {et~al.}(2015)\citenamefont
  {van Donkelaar}, \citenamefont {Yang}, \citenamefont {Alves}, \citenamefont
  {McCallum}, \citenamefont {Hougaard}, \citenamefont {Johnson}, \citenamefont
  {Hudson}, \citenamefont {Dzurak}, \citenamefont {Morello}, \citenamefont
  {Spemann},\ and\ \citenamefont {Jamieson}}]{DYA+15}%
  \BibitemOpen
  \bibfield  {author} {\bibinfo {author} {\bibfnamefont {J.}~\bibnamefont {van
  Donkelaar}}, \bibinfo {author} {\bibfnamefont {C.}~\bibnamefont {Yang}},
  \bibinfo {author} {\bibfnamefont {A.~D.~C.}\ \bibnamefont {Alves}}, \bibinfo
  {author} {\bibfnamefont {J.~C.}\ \bibnamefont {McCallum}}, \bibinfo {author}
  {\bibfnamefont {C.}~\bibnamefont {Hougaard}}, \bibinfo {author}
  {\bibfnamefont {B.~C.}\ \bibnamefont {Johnson}}, \bibinfo {author}
  {\bibfnamefont {F.~E.}\ \bibnamefont {Hudson}}, \bibinfo {author}
  {\bibfnamefont {A.~S.}\ \bibnamefont {Dzurak}}, \bibinfo {author}
  {\bibfnamefont {A.}~\bibnamefont {Morello}}, \bibinfo {author} {\bibfnamefont
  {D.}~\bibnamefont {Spemann}}, \ and\ \bibinfo {author} {\bibfnamefont
  {D.~N.}\ \bibnamefont {Jamieson}},\ }\href@noop {} {\bibfield  {journal}
  {\bibinfo  {journal} {J.\ Phys.: Condens.\ Matter}\ }\textbf {\bibinfo
  {volume} {27}},\ \bibinfo {pages} {154204} (\bibinfo {year}
  {2015})}\BibitemShut {NoStop}%
\bibitem [{\citenamefont {Togan}\ \emph {et~al.}(2010)\citenamefont {Togan},
  \citenamefont {Chu}, \citenamefont {Trifonov}, \citenamefont {Jiang},
  \citenamefont {Maze}, \citenamefont {Childress}, \citenamefont {Dutt},
  \citenamefont {S{\o}rensen}, \citenamefont {Hemmer}, \citenamefont {Zibrov},\
  and\ \citenamefont {Lukin}}]{TCT+10}%
  \BibitemOpen
  \bibfield  {author} {\bibinfo {author} {\bibfnamefont {E.}~\bibnamefont
  {Togan}}, \bibinfo {author} {\bibfnamefont {Y.}~\bibnamefont {Chu}}, \bibinfo
  {author} {\bibfnamefont {A.~S.}\ \bibnamefont {Trifonov}}, \bibinfo {author}
  {\bibfnamefont {L.}~\bibnamefont {Jiang}}, \bibinfo {author} {\bibfnamefont
  {J.}~\bibnamefont {Maze}}, \bibinfo {author} {\bibfnamefont {L.}~\bibnamefont
  {Childress}}, \bibinfo {author} {\bibfnamefont {M.~V.~G.}\ \bibnamefont
  {Dutt}}, \bibinfo {author} {\bibfnamefont {A.~S.}\ \bibnamefont
  {S{\o}rensen}}, \bibinfo {author} {\bibfnamefont {P.~R.}\ \bibnamefont
  {Hemmer}}, \bibinfo {author} {\bibfnamefont {A.~S.}\ \bibnamefont {Zibrov}},
  \ and\ \bibinfo {author} {\bibfnamefont {M.~D.}\ \bibnamefont {Lukin}},\
  }\href@noop {} {\bibfield  {journal} {\bibinfo  {journal} {Nature}\ }\textbf
  {\bibinfo {volume} {466}},\ \bibinfo {pages} {730} (\bibinfo {year}
  {2010})}\BibitemShut {NoStop}%
\bibitem [{\citenamefont {Bernien}\ \emph {et~al.}(2013)\citenamefont
  {Bernien}, \citenamefont {Hensen}, \citenamefont {Pfaff}, \citenamefont
  {Koolstra}, \citenamefont {Blok}, \citenamefont {Robledo}, \citenamefont
  {Taminiau}, \citenamefont {Markham}, \citenamefont {Twitchen}, \citenamefont
  {Childress},\ and\ \citenamefont {Hanson}}]{BHP+13}%
  \BibitemOpen
  \bibfield  {author} {\bibinfo {author} {\bibfnamefont {H.}~\bibnamefont
  {Bernien}}, \bibinfo {author} {\bibfnamefont {B.}~\bibnamefont {Hensen}},
  \bibinfo {author} {\bibfnamefont {W.}~\bibnamefont {Pfaff}}, \bibinfo
  {author} {\bibfnamefont {G.}~\bibnamefont {Koolstra}}, \bibinfo {author}
  {\bibfnamefont {M.~S.}\ \bibnamefont {Blok}}, \bibinfo {author}
  {\bibfnamefont {L.}~\bibnamefont {Robledo}}, \bibinfo {author} {\bibfnamefont
  {T.~H.}\ \bibnamefont {Taminiau}}, \bibinfo {author} {\bibfnamefont
  {M.}~\bibnamefont {Markham}}, \bibinfo {author} {\bibfnamefont {D.~J.}\
  \bibnamefont {Twitchen}}, \bibinfo {author} {\bibfnamefont {L.}~\bibnamefont
  {Childress}}, \ and\ \bibinfo {author} {\bibfnamefont {R.}~\bibnamefont
  {Hanson}},\ }\href@noop {} {\bibfield  {journal} {\bibinfo  {journal}
  {Nature}\ }\textbf {\bibinfo {volume} {497}},\ \bibinfo {pages} {86}
  (\bibinfo {year} {2013})}\BibitemShut {NoStop}%
\bibitem [{\citenamefont {Pfaff}\ \emph {et~al.}(2014)\citenamefont {Pfaff},
  \citenamefont {Hensen}, \citenamefont {Bernien}, \citenamefont {van Dam},
  \citenamefont {Blok}, \citenamefont {Taminiau}, \citenamefont {Tiggelman},
  \citenamefont {Schouten}, \citenamefont {Markham}, \citenamefont {Twitchen},\
  and\ \citenamefont {Hanson}}]{PHB+14}%
  \BibitemOpen
  \bibfield  {author} {\bibinfo {author} {\bibfnamefont {W.}~\bibnamefont
  {Pfaff}}, \bibinfo {author} {\bibfnamefont {B.~J.}\ \bibnamefont {Hensen}},
  \bibinfo {author} {\bibfnamefont {H.}~\bibnamefont {Bernien}}, \bibinfo
  {author} {\bibfnamefont {S.~B.}\ \bibnamefont {van Dam}}, \bibinfo {author}
  {\bibfnamefont {M.~S.}\ \bibnamefont {Blok}}, \bibinfo {author}
  {\bibfnamefont {T.~H.}\ \bibnamefont {Taminiau}}, \bibinfo {author}
  {\bibfnamefont {M.~J.}\ \bibnamefont {Tiggelman}}, \bibinfo {author}
  {\bibfnamefont {R.~N.}\ \bibnamefont {Schouten}}, \bibinfo {author}
  {\bibfnamefont {M.}~\bibnamefont {Markham}}, \bibinfo {author} {\bibfnamefont
  {D.~J.}\ \bibnamefont {Twitchen}}, \ and\ \bibinfo {author} {\bibfnamefont
  {R.}~\bibnamefont {Hanson}},\ }\href@noop {} {\bibfield  {journal} {\bibinfo
  {journal} {Science}\ }\textbf {\bibinfo {volume} {345}},\ \bibinfo {pages}
  {532} (\bibinfo {year} {2014})}\BibitemShut {NoStop}%
\bibitem [{\citenamefont {Hensen}\ \emph {et~al.}(2015)\citenamefont {Hensen},
  \citenamefont {Bernien}, \citenamefont {Dr{\'e}au}, \citenamefont {Reiserer},
  \citenamefont {Kalb}, \citenamefont {Blok}, \citenamefont {Ruitenberg},
  \citenamefont {Vermeulen}, \citenamefont {Schouten}, \citenamefont
  {Abell{\'a}n}, \citenamefont {Amaya}, \citenamefont {Pruneri}, \citenamefont
  {Mitchell}, \citenamefont {Markham}, \citenamefont {Twitchen}, \citenamefont
  {Elkouss}, \citenamefont {Wehner}, \citenamefont {Taminiau},\ and\
  \citenamefont {Hanson}}]{HBD+15}%
  \BibitemOpen
  \bibfield  {author} {\bibinfo {author} {\bibfnamefont {B.}~\bibnamefont
  {Hensen}}, \bibinfo {author} {\bibfnamefont {H.}~\bibnamefont {Bernien}},
  \bibinfo {author} {\bibfnamefont {A.~E.}\ \bibnamefont {Dr{\'e}au}}, \bibinfo
  {author} {\bibfnamefont {A.}~\bibnamefont {Reiserer}}, \bibinfo {author}
  {\bibfnamefont {N.}~\bibnamefont {Kalb}}, \bibinfo {author} {\bibfnamefont
  {M.~S.}\ \bibnamefont {Blok}}, \bibinfo {author} {\bibfnamefont
  {J.}~\bibnamefont {Ruitenberg}}, \bibinfo {author} {\bibfnamefont {R.~F.~L.}\
  \bibnamefont {Vermeulen}}, \bibinfo {author} {\bibfnamefont {R.~N.}\
  \bibnamefont {Schouten}}, \bibinfo {author} {\bibfnamefont {C.}~\bibnamefont
  {Abell{\'a}n}}, \bibinfo {author} {\bibfnamefont {W.}~\bibnamefont {Amaya}},
  \bibinfo {author} {\bibfnamefont {V.}~\bibnamefont {Pruneri}}, \bibinfo
  {author} {\bibfnamefont {M.~W.}\ \bibnamefont {Mitchell}}, \bibinfo {author}
  {\bibfnamefont {M.}~\bibnamefont {Markham}}, \bibinfo {author} {\bibfnamefont
  {D.~J.}\ \bibnamefont {Twitchen}}, \bibinfo {author} {\bibfnamefont
  {D.}~\bibnamefont {Elkouss}}, \bibinfo {author} {\bibfnamefont
  {S.}~\bibnamefont {Wehner}}, \bibinfo {author} {\bibfnamefont {T.~H.}\
  \bibnamefont {Taminiau}}, \ and\ \bibinfo {author} {\bibfnamefont
  {R.}~\bibnamefont {Hanson}},\ }\href@noop {} {\bibfield  {journal} {\bibinfo
  {journal} {Nature}\ }\textbf {\bibinfo {volume} {526}},\ \bibinfo {pages}
  {682} (\bibinfo {year} {2015})}\BibitemShut {NoStop}%
\bibitem [{\citenamefont {Maze}\ \emph {et~al.}(2008)\citenamefont {Maze},
  \citenamefont {Stanwix}, \citenamefont {Hodges}, \citenamefont {Hong},
  \citenamefont {Taylor}, \citenamefont {Cappellaro}, \citenamefont {Jiang},
  \citenamefont {Dutt}, \citenamefont {Togan}, \citenamefont {Zibrov},
  \citenamefont {Yacoby}, \citenamefont {Walsworth},\ and\ \citenamefont
  {Lukin}}]{MSH+08}%
  \BibitemOpen
  \bibfield  {author} {\bibinfo {author} {\bibfnamefont {J.~R.}\ \bibnamefont
  {Maze}}, \bibinfo {author} {\bibfnamefont {P.~L.}\ \bibnamefont {Stanwix}},
  \bibinfo {author} {\bibfnamefont {J.~S.}\ \bibnamefont {Hodges}}, \bibinfo
  {author} {\bibfnamefont {S.}~\bibnamefont {Hong}}, \bibinfo {author}
  {\bibfnamefont {J.~M.}\ \bibnamefont {Taylor}}, \bibinfo {author}
  {\bibfnamefont {P.}~\bibnamefont {Cappellaro}}, \bibinfo {author}
  {\bibfnamefont {L.}~\bibnamefont {Jiang}}, \bibinfo {author} {\bibfnamefont
  {M.~V.~G.}\ \bibnamefont {Dutt}}, \bibinfo {author} {\bibfnamefont
  {E.}~\bibnamefont {Togan}}, \bibinfo {author} {\bibfnamefont {A.~S.}\
  \bibnamefont {Zibrov}}, \bibinfo {author} {\bibfnamefont {A.}~\bibnamefont
  {Yacoby}}, \bibinfo {author} {\bibfnamefont {R.~L.}\ \bibnamefont
  {Walsworth}}, \ and\ \bibinfo {author} {\bibfnamefont {M.~D.}\ \bibnamefont
  {Lukin}},\ }\href@noop {} {\bibfield  {journal} {\bibinfo  {journal}
  {Nature}\ }\textbf {\bibinfo {volume} {455}},\ \bibinfo {pages} {644}
  (\bibinfo {year} {2008})}\BibitemShut {NoStop}%
\bibitem [{\citenamefont {Balasubramanian}\ \emph {et~al.}(2008)\citenamefont
  {Balasubramanian}, \citenamefont {Chan}, \citenamefont {Kolesov},
  \citenamefont {Al-Hmoud}, \citenamefont {Tisler}, \citenamefont {Shin},
  \citenamefont {Kim}, \citenamefont {Wojcik}, \citenamefont {Hemmer},
  \citenamefont {Krueger}, \citenamefont {Hanke}, \citenamefont
  {Leitenstorfer}, \citenamefont {Bratschitsch}, \citenamefont {Jelezko},\ and\
  \citenamefont {Wrachtrup}}]{BCK+08}%
  \BibitemOpen
  \bibfield  {author} {\bibinfo {author} {\bibfnamefont {G.}~\bibnamefont
  {Balasubramanian}}, \bibinfo {author} {\bibfnamefont {I.~Y.}\ \bibnamefont
  {Chan}}, \bibinfo {author} {\bibfnamefont {R.}~\bibnamefont {Kolesov}},
  \bibinfo {author} {\bibfnamefont {M.}~\bibnamefont {Al-Hmoud}}, \bibinfo
  {author} {\bibfnamefont {J.}~\bibnamefont {Tisler}}, \bibinfo {author}
  {\bibfnamefont {C.}~\bibnamefont {Shin}}, \bibinfo {author} {\bibfnamefont
  {C.}~\bibnamefont {Kim}}, \bibinfo {author} {\bibfnamefont {A.}~\bibnamefont
  {Wojcik}}, \bibinfo {author} {\bibfnamefont {P.~R.}\ \bibnamefont {Hemmer}},
  \bibinfo {author} {\bibfnamefont {A.}~\bibnamefont {Krueger}}, \bibinfo
  {author} {\bibfnamefont {T.}~\bibnamefont {Hanke}}, \bibinfo {author}
  {\bibfnamefont {A.}~\bibnamefont {Leitenstorfer}}, \bibinfo {author}
  {\bibfnamefont {R.}~\bibnamefont {Bratschitsch}}, \bibinfo {author}
  {\bibfnamefont {F.}~\bibnamefont {Jelezko}}, \ and\ \bibinfo {author}
  {\bibfnamefont {J.}~\bibnamefont {Wrachtrup}},\ }\href@noop {} {\bibfield
  {journal} {\bibinfo  {journal} {Nature}\ }\textbf {\bibinfo {volume} {455}},\
  \bibinfo {pages} {648} (\bibinfo {year} {2008})}\BibitemShut {NoStop}%
\bibitem [{\citenamefont {Rondin}\ \emph {et~al.}(2014)\citenamefont {Rondin},
  \citenamefont {Tetienne}, \citenamefont {Hingant}, \citenamefont {Roch},
  \citenamefont {Maletinsky},\ and\ \citenamefont {Jacques}}]{RTH+14}%
  \BibitemOpen
  \bibfield  {author} {\bibinfo {author} {\bibfnamefont {L.}~\bibnamefont
  {Rondin}}, \bibinfo {author} {\bibfnamefont {J.-P.}\ \bibnamefont
  {Tetienne}}, \bibinfo {author} {\bibfnamefont {T.}~\bibnamefont {Hingant}},
  \bibinfo {author} {\bibfnamefont {J.-F.}\ \bibnamefont {Roch}}, \bibinfo
  {author} {\bibfnamefont {P.}~\bibnamefont {Maletinsky}}, \ and\ \bibinfo
  {author} {\bibfnamefont {V.}~\bibnamefont {Jacques}},\ }\href@noop {}
  {\bibfield  {journal} {\bibinfo  {journal} {Rep.\ Prog.\ Phys.}\ }\textbf
  {\bibinfo {volume} {77}},\ \bibinfo {pages} {056503} (\bibinfo {year}
  {2014})}\BibitemShut {NoStop}%
\bibitem [{\citenamefont {Wrachtrup}\ \emph {et~al.}(2013)\citenamefont
  {Wrachtrup}, \citenamefont {Jelezko}, \citenamefont {Grotz},\ and\
  \citenamefont {McGuinness}}]{WJGM13}%
  \BibitemOpen
  \bibfield  {author} {\bibinfo {author} {\bibfnamefont {J.}~\bibnamefont
  {Wrachtrup}}, \bibinfo {author} {\bibfnamefont {F.}~\bibnamefont {Jelezko}},
  \bibinfo {author} {\bibfnamefont {B.}~\bibnamefont {Grotz}}, \ and\ \bibinfo
  {author} {\bibfnamefont {L.}~\bibnamefont {McGuinness}},\ }\href@noop {}
  {\bibfield  {journal} {\bibinfo  {journal} {MRS Bull.}\ }\textbf {\bibinfo
  {volume} {38}},\ \bibinfo {pages} {149} (\bibinfo {year} {2013})}\BibitemShut
  {NoStop}%
\bibitem [{\citenamefont {Schr{\"o}der}\ \emph {et~al.}(2016)\citenamefont
  {Schr{\"o}der}, \citenamefont {Mouradian}, \citenamefont {Zheng},
  \citenamefont {Trusheim}, \citenamefont {Walsh}, \citenamefont {Chen},
  \citenamefont {Li}, \citenamefont {Bayn},\ and\ \citenamefont
  {Englund}}]{SMZ+16}%
  \BibitemOpen
  \bibfield  {author} {\bibinfo {author} {\bibfnamefont {T.}~\bibnamefont
  {Schr{\"o}der}}, \bibinfo {author} {\bibfnamefont {S.~L.}\ \bibnamefont
  {Mouradian}}, \bibinfo {author} {\bibfnamefont {J.}~\bibnamefont {Zheng}},
  \bibinfo {author} {\bibfnamefont {M.~E.}\ \bibnamefont {Trusheim}}, \bibinfo
  {author} {\bibfnamefont {M.}~\bibnamefont {Walsh}}, \bibinfo {author}
  {\bibfnamefont {E.~H.}\ \bibnamefont {Chen}}, \bibinfo {author}
  {\bibfnamefont {L.}~\bibnamefont {Li}}, \bibinfo {author} {\bibfnamefont
  {I.}~\bibnamefont {Bayn}}, \ and\ \bibinfo {author} {\bibfnamefont
  {D.}~\bibnamefont {Englund}},\ }\href@noop {} {\bibfield  {journal} {\bibinfo
   {journal} {J.\ Opt.\ Soc.\ Am.\ B}\ }\textbf {\bibinfo {volume} {33}},\
  \bibinfo {pages} {B65} (\bibinfo {year} {2016})}\BibitemShut {NoStop}%
\bibitem [{\citenamefont {Ohno}\ \emph {et~al.}(2012)\citenamefont {Ohno},
  \citenamefont {Heremans}, \citenamefont {Bassett}, \citenamefont {Myers},
  \citenamefont {Toyli}, \citenamefont {Bleszynski~Jayich}, \citenamefont
  {Palmstr{\o}m},\ and\ \citenamefont {Awschalom}}]{OHB+12}%
  \BibitemOpen
  \bibfield  {author} {\bibinfo {author} {\bibfnamefont {K.}~\bibnamefont
  {Ohno}}, \bibinfo {author} {\bibfnamefont {F.~J.}\ \bibnamefont {Heremans}},
  \bibinfo {author} {\bibfnamefont {L.~C.}\ \bibnamefont {Bassett}}, \bibinfo
  {author} {\bibfnamefont {B.~A.}\ \bibnamefont {Myers}}, \bibinfo {author}
  {\bibfnamefont {D.~M.}\ \bibnamefont {Toyli}}, \bibinfo {author}
  {\bibfnamefont {A.~C.}\ \bibnamefont {Bleszynski~Jayich}}, \bibinfo {author}
  {\bibfnamefont {C.~J.}\ \bibnamefont {Palmstr{\o}m}}, \ and\ \bibinfo
  {author} {\bibfnamefont {D.~D.}\ \bibnamefont {Awschalom}},\ }\href@noop {}
  {\bibfield  {journal} {\bibinfo  {journal} {Appl.\ Phys.\ Lett.}\ }\textbf
  {\bibinfo {volume} {101}},\ \bibinfo {pages} {082413} (\bibinfo {year}
  {2012})}\BibitemShut {NoStop}%
\bibitem [{\citenamefont {Ohashi}\ \emph {et~al.}(2013)\citenamefont {Ohashi},
  \citenamefont {Rosskopf}, \citenamefont {Watanabe}, \citenamefont {Loretz},
  \citenamefont {Tao}, \citenamefont {Hauert}, \citenamefont {Tomizawa},
  \citenamefont {Ishikawa}, \citenamefont {Ishi-Hayase}, \citenamefont
  {Shikata}, \citenamefont {Degen},\ and\ \citenamefont {Itoh}}]{ORW+13}%
  \BibitemOpen
  \bibfield  {author} {\bibinfo {author} {\bibfnamefont {K.}~\bibnamefont
  {Ohashi}}, \bibinfo {author} {\bibfnamefont {T.}~\bibnamefont {Rosskopf}},
  \bibinfo {author} {\bibfnamefont {H.}~\bibnamefont {Watanabe}}, \bibinfo
  {author} {\bibfnamefont {M.}~\bibnamefont {Loretz}}, \bibinfo {author}
  {\bibfnamefont {Y.}~\bibnamefont {Tao}}, \bibinfo {author} {\bibfnamefont
  {R.}~\bibnamefont {Hauert}}, \bibinfo {author} {\bibfnamefont
  {S.}~\bibnamefont {Tomizawa}}, \bibinfo {author} {\bibfnamefont
  {T.}~\bibnamefont {Ishikawa}}, \bibinfo {author} {\bibfnamefont
  {J.}~\bibnamefont {Ishi-Hayase}}, \bibinfo {author} {\bibfnamefont
  {S.}~\bibnamefont {Shikata}}, \bibinfo {author} {\bibfnamefont {C.~L.}\
  \bibnamefont {Degen}}, \ and\ \bibinfo {author} {\bibfnamefont {K.~M.}\
  \bibnamefont {Itoh}},\ }\href@noop {} {\bibfield  {journal} {\bibinfo
  {journal} {Nano Lett.}\ }\textbf {\bibinfo {volume} {13}},\ \bibinfo {pages}
  {4733} (\bibinfo {year} {2013})}\BibitemShut {NoStop}%
\bibitem [{\citenamefont {Ohno}\ \emph {et~al.}(2014)\citenamefont {Ohno},
  \citenamefont {Heremans}, \citenamefont {de~las Casas}, \citenamefont
  {Myers}, \citenamefont {Alem{\'a}n}, \citenamefont {Bleszynski~Jayich},\ and\
  \citenamefont {Awschalom}}]{OHB+14}%
  \BibitemOpen
  \bibfield  {author} {\bibinfo {author} {\bibfnamefont {K.}~\bibnamefont
  {Ohno}}, \bibinfo {author} {\bibfnamefont {F.~J.}\ \bibnamefont {Heremans}},
  \bibinfo {author} {\bibfnamefont {C.~F.}\ \bibnamefont {de~las Casas}},
  \bibinfo {author} {\bibfnamefont {B.~A.}\ \bibnamefont {Myers}}, \bibinfo
  {author} {\bibfnamefont {B.~J.}\ \bibnamefont {Alem{\'a}n}}, \bibinfo
  {author} {\bibfnamefont {A.~C.}\ \bibnamefont {Bleszynski~Jayich}}, \ and\
  \bibinfo {author} {\bibfnamefont {D.~D.}\ \bibnamefont {Awschalom}},\
  }\href@noop {} {\bibfield  {journal} {\bibinfo  {journal} {Appl.\ Phys.\
  Lett.}\ }\textbf {\bibinfo {volume} {105}},\ \bibinfo {pages} {052406}
  (\bibinfo {year} {2014})}\BibitemShut {NoStop}%
\bibitem [{\citenamefont {Kleinsasser}\ \emph {et~al.}(2016)\citenamefont
  {Kleinsasser}, \citenamefont {Stanfield}, \citenamefont {Banks},
  \citenamefont {Zhu}, \citenamefont {Li}, \citenamefont {Acosta},
  \citenamefont {Watanabe}, \citenamefont {Itoh},\ and\ \citenamefont
  {Fu}}]{KSB+16}%
  \BibitemOpen
  \bibfield  {author} {\bibinfo {author} {\bibfnamefont {E.~E.}\ \bibnamefont
  {Kleinsasser}}, \bibinfo {author} {\bibfnamefont {M.~M.}\ \bibnamefont
  {Stanfield}}, \bibinfo {author} {\bibfnamefont {J.~K.~Q.}\ \bibnamefont
  {Banks}}, \bibinfo {author} {\bibfnamefont {Z.}~\bibnamefont {Zhu}}, \bibinfo
  {author} {\bibfnamefont {W.-D.}\ \bibnamefont {Li}}, \bibinfo {author}
  {\bibfnamefont {V.~M.}\ \bibnamefont {Acosta}}, \bibinfo {author}
  {\bibfnamefont {H.}~\bibnamefont {Watanabe}}, \bibinfo {author}
  {\bibfnamefont {K.~M.}\ \bibnamefont {Itoh}}, \ and\ \bibinfo {author}
  {\bibfnamefont {K.-M.~C.}\ \bibnamefont {Fu}},\ }\href@noop {} {\bibfield
  {journal} {\bibinfo  {journal} {Appl.\ Phys.\ Lett.}\ }\textbf {\bibinfo
  {volume} {108}},\ \bibinfo {pages} {202401} (\bibinfo {year}
  {2016})}\BibitemShut {NoStop}%
\bibitem [{\citenamefont {Meijer}\ \emph {et~al.}(2005)\citenamefont {Meijer},
  \citenamefont {Burchard}, \citenamefont {Domhan}, \citenamefont {Wittmann},
  \citenamefont {Gaebel}, \citenamefont {Popa}, \citenamefont {Jelezko},\ and\
  \citenamefont {Wrachtrup}}]{MBD+05}%
  \BibitemOpen
  \bibfield  {author} {\bibinfo {author} {\bibfnamefont {J.}~\bibnamefont
  {Meijer}}, \bibinfo {author} {\bibfnamefont {B.}~\bibnamefont {Burchard}},
  \bibinfo {author} {\bibfnamefont {M.}~\bibnamefont {Domhan}}, \bibinfo
  {author} {\bibfnamefont {C.}~\bibnamefont {Wittmann}}, \bibinfo {author}
  {\bibfnamefont {T.}~\bibnamefont {Gaebel}}, \bibinfo {author} {\bibfnamefont
  {I.}~\bibnamefont {Popa}}, \bibinfo {author} {\bibfnamefont {F.}~\bibnamefont
  {Jelezko}}, \ and\ \bibinfo {author} {\bibfnamefont {J.}~\bibnamefont
  {Wrachtrup}},\ }\href@noop {} {\bibfield  {journal} {\bibinfo  {journal}
  {Appl.\ Phys.\ Lett.}\ }\textbf {\bibinfo {volume} {87}},\ \bibinfo {pages}
  {261909} (\bibinfo {year} {2005})}\BibitemShut {NoStop}%
\bibitem [{\citenamefont {Rabeau}\ \emph {et~al.}(2006)\citenamefont {Rabeau},
  \citenamefont {Reichart}, \citenamefont {Tamanyan}, \citenamefont {Jamieson},
  \citenamefont {Prawer}, \citenamefont {Jelezko}, \citenamefont {Gaebel},
  \citenamefont {Popa}, \citenamefont {Domhan},\ and\ \citenamefont
  {Wrachtrup}}]{RRT+06}%
  \BibitemOpen
  \bibfield  {author} {\bibinfo {author} {\bibfnamefont {J.~R.}\ \bibnamefont
  {Rabeau}}, \bibinfo {author} {\bibfnamefont {P.}~\bibnamefont {Reichart}},
  \bibinfo {author} {\bibfnamefont {G.}~\bibnamefont {Tamanyan}}, \bibinfo
  {author} {\bibfnamefont {D.~N.}\ \bibnamefont {Jamieson}}, \bibinfo {author}
  {\bibfnamefont {S.}~\bibnamefont {Prawer}}, \bibinfo {author} {\bibfnamefont
  {F.}~\bibnamefont {Jelezko}}, \bibinfo {author} {\bibfnamefont
  {T.}~\bibnamefont {Gaebel}}, \bibinfo {author} {\bibfnamefont
  {I.}~\bibnamefont {Popa}}, \bibinfo {author} {\bibfnamefont {M.}~\bibnamefont
  {Domhan}}, \ and\ \bibinfo {author} {\bibfnamefont {J.}~\bibnamefont
  {Wrachtrup}},\ }\href@noop {} {\bibfield  {journal} {\bibinfo  {journal}
  {Appl.\ Phys.\ Lett.}\ }\textbf {\bibinfo {volume} {88}},\ \bibinfo {pages}
  {023113} (\bibinfo {year} {2006})}\BibitemShut {NoStop}%
\bibitem [{\citenamefont {Toyli}\ \emph {et~al.}(2010)\citenamefont {Toyli},
  \citenamefont {Weis}, \citenamefont {Fuchs}, \citenamefont {Schenkel},\ and\
  \citenamefont {Awschalom}}]{TWF+10}%
  \BibitemOpen
  \bibfield  {author} {\bibinfo {author} {\bibfnamefont {D.~M.}\ \bibnamefont
  {Toyli}}, \bibinfo {author} {\bibfnamefont {C.~D.}\ \bibnamefont {Weis}},
  \bibinfo {author} {\bibfnamefont {G.~D.}\ \bibnamefont {Fuchs}}, \bibinfo
  {author} {\bibfnamefont {T.}~\bibnamefont {Schenkel}}, \ and\ \bibinfo
  {author} {\bibfnamefont {D.~D.}\ \bibnamefont {Awschalom}},\ }\href@noop {}
  {\bibfield  {journal} {\bibinfo  {journal} {Nano Lett.}\ }\textbf {\bibinfo
  {volume} {10}},\ \bibinfo {pages} {3168} (\bibinfo {year}
  {2010})}\BibitemShut {NoStop}%
\bibitem [{\citenamefont {Spinicelli}\ \emph {et~al.}(2011)\citenamefont
  {Spinicelli}, \citenamefont {Dr{\'e}au}, \citenamefont {Rondin},
  \citenamefont {Silva}, \citenamefont {Achard}, \citenamefont {Xavier},
  \citenamefont {Bansropun}, \citenamefont {Debuisschert}, \citenamefont
  {Pezzagna},\ and\ \citenamefont {Meijer}}]{SDR+11}%
  \BibitemOpen
  \bibfield  {author} {\bibinfo {author} {\bibfnamefont {P.}~\bibnamefont
  {Spinicelli}}, \bibinfo {author} {\bibfnamefont {A.}~\bibnamefont
  {Dr{\'e}au}}, \bibinfo {author} {\bibfnamefont {L.}~\bibnamefont {Rondin}},
  \bibinfo {author} {\bibfnamefont {F.}~\bibnamefont {Silva}}, \bibinfo
  {author} {\bibfnamefont {J.}~\bibnamefont {Achard}}, \bibinfo {author}
  {\bibfnamefont {S.}~\bibnamefont {Xavier}}, \bibinfo {author} {\bibfnamefont
  {S.}~\bibnamefont {Bansropun}}, \bibinfo {author} {\bibfnamefont
  {T.}~\bibnamefont {Debuisschert}}, \bibinfo {author} {\bibfnamefont
  {S.}~\bibnamefont {Pezzagna}}, \ and\ \bibinfo {author} {\bibfnamefont
  {J.}~\bibnamefont {Meijer}},\ }\href@noop {} {\bibfield  {journal} {\bibinfo
  {journal} {New J.\ Phys.}\ }\textbf {\bibinfo {volume} {13}},\ \bibinfo
  {pages} {025014} (\bibinfo {year} {2011})}\BibitemShut {NoStop}%
\bibitem [{\citenamefont {Pezzagna}\ \emph {et~al.}(2011)\citenamefont
  {Pezzagna}, \citenamefont {Rogalla}, \citenamefont {Becker}, \citenamefont
  {Jakobi}, \citenamefont {Dolde}, \citenamefont {Naydenov}, \citenamefont
  {Wrachtrup}, \citenamefont {Jelezko}, \citenamefont {Trautmann},\ and\
  \citenamefont {Meijer}}]{PRB+11}%
  \BibitemOpen
  \bibfield  {author} {\bibinfo {author} {\bibfnamefont {S.}~\bibnamefont
  {Pezzagna}}, \bibinfo {author} {\bibfnamefont {D.}~\bibnamefont {Rogalla}},
  \bibinfo {author} {\bibfnamefont {H.-W.}\ \bibnamefont {Becker}}, \bibinfo
  {author} {\bibfnamefont {I.}~\bibnamefont {Jakobi}}, \bibinfo {author}
  {\bibfnamefont {F.}~\bibnamefont {Dolde}}, \bibinfo {author} {\bibfnamefont
  {B.}~\bibnamefont {Naydenov}}, \bibinfo {author} {\bibfnamefont
  {J.}~\bibnamefont {Wrachtrup}}, \bibinfo {author} {\bibfnamefont
  {F.}~\bibnamefont {Jelezko}}, \bibinfo {author} {\bibfnamefont
  {C.}~\bibnamefont {Trautmann}}, \ and\ \bibinfo {author} {\bibfnamefont
  {J.}~\bibnamefont {Meijer}},\ }\href@noop {} {\bibfield  {journal} {\bibinfo
  {journal} {Phys. Status Solidi A}\ }\textbf {\bibinfo {volume} {208}},\
  \bibinfo {pages} {2017} (\bibinfo {year} {2011})}\BibitemShut {NoStop}%
\bibitem [{\citenamefont {Sangtawesin}\ \emph {et~al.}(2014)\citenamefont
  {Sangtawesin}, \citenamefont {Brundage}, \citenamefont {Atkins},\ and\
  \citenamefont {Petta}}]{SBAP14}%
  \BibitemOpen
  \bibfield  {author} {\bibinfo {author} {\bibfnamefont {S.}~\bibnamefont
  {Sangtawesin}}, \bibinfo {author} {\bibfnamefont {T.~O.}\ \bibnamefont
  {Brundage}}, \bibinfo {author} {\bibfnamefont {Z.~J.}\ \bibnamefont
  {Atkins}}, \ and\ \bibinfo {author} {\bibfnamefont {J.~R.}\ \bibnamefont
  {Petta}},\ }\href@noop {} {\bibfield  {journal} {\bibinfo  {journal} {Appl.\
  Phys.\ Lett.}\ }\textbf {\bibinfo {volume} {105}},\ \bibinfo {pages} {063107}
  (\bibinfo {year} {2014})}\BibitemShut {NoStop}%
\bibitem [{\citenamefont {Pezzagna}\ \emph {et~al.}(2010)\citenamefont
  {Pezzagna}, \citenamefont {Naydenov}, \citenamefont {Jelezko}, \citenamefont
  {Wrachtrup},\ and\ \citenamefont {Meijer}}]{PNJ+10}%
  \BibitemOpen
  \bibfield  {author} {\bibinfo {author} {\bibfnamefont {S.}~\bibnamefont
  {Pezzagna}}, \bibinfo {author} {\bibfnamefont {B.}~\bibnamefont {Naydenov}},
  \bibinfo {author} {\bibfnamefont {F.}~\bibnamefont {Jelezko}}, \bibinfo
  {author} {\bibfnamefont {J.}~\bibnamefont {Wrachtrup}}, \ and\ \bibinfo
  {author} {\bibfnamefont {J.}~\bibnamefont {Meijer}},\ }\href@noop {}
  {\bibfield  {journal} {\bibinfo  {journal} {New J.\ Phys.}\ }\textbf
  {\bibinfo {volume} {12}},\ \bibinfo {pages} {065017} (\bibinfo {year}
  {2010})}\BibitemShut {NoStop}%
\bibitem [{\citenamefont {Ofori-Okai}\ \emph {et~al.}(2012)\citenamefont
  {Ofori-Okai}, \citenamefont {Pezzagna}, \citenamefont {Chang}, \citenamefont
  {Loretz}, \citenamefont {Schirhagl}, \citenamefont {Tao}, \citenamefont
  {Moores}, \citenamefont {Groot-Berning}, \citenamefont {Meijer},\ and\
  \citenamefont {Degen}}]{OPC+12}%
  \BibitemOpen
  \bibfield  {author} {\bibinfo {author} {\bibfnamefont {B.~K.}\ \bibnamefont
  {Ofori-Okai}}, \bibinfo {author} {\bibfnamefont {S.}~\bibnamefont
  {Pezzagna}}, \bibinfo {author} {\bibfnamefont {K.}~\bibnamefont {Chang}},
  \bibinfo {author} {\bibfnamefont {M.}~\bibnamefont {Loretz}}, \bibinfo
  {author} {\bibfnamefont {R.}~\bibnamefont {Schirhagl}}, \bibinfo {author}
  {\bibfnamefont {Y.}~\bibnamefont {Tao}}, \bibinfo {author} {\bibfnamefont
  {B.~A.}\ \bibnamefont {Moores}}, \bibinfo {author} {\bibfnamefont
  {K.}~\bibnamefont {Groot-Berning}}, \bibinfo {author} {\bibfnamefont
  {J.}~\bibnamefont {Meijer}}, \ and\ \bibinfo {author} {\bibfnamefont {C.~L.}\
  \bibnamefont {Degen}},\ }\href@noop {} {\bibfield  {journal} {\bibinfo
  {journal} {Phys.\ Rev.\ B}\ }\textbf {\bibinfo {volume} {86}},\ \bibinfo
  {pages} {081406} (\bibinfo {year} {2012})}\BibitemShut {NoStop}%
\bibitem [{\citenamefont {Osterkamp}\ \emph {et~al.}(2013)\citenamefont
  {Osterkamp}, \citenamefont {Scharpf}, \citenamefont {Pezzagna}, \citenamefont
  {Meijer}, \citenamefont {Diemant}, \citenamefont {Behm}, \citenamefont
  {Naydenov},\ and\ \citenamefont {Jelezko}}]{OSP+13}%
  \BibitemOpen
  \bibfield  {author} {\bibinfo {author} {\bibfnamefont {C.}~\bibnamefont
  {Osterkamp}}, \bibinfo {author} {\bibfnamefont {J.}~\bibnamefont {Scharpf}},
  \bibinfo {author} {\bibfnamefont {S.}~\bibnamefont {Pezzagna}}, \bibinfo
  {author} {\bibfnamefont {J.}~\bibnamefont {Meijer}}, \bibinfo {author}
  {\bibfnamefont {T.}~\bibnamefont {Diemant}}, \bibinfo {author} {\bibfnamefont
  {R.~J.}\ \bibnamefont {Behm}}, \bibinfo {author} {\bibfnamefont
  {B.}~\bibnamefont {Naydenov}}, \ and\ \bibinfo {author} {\bibfnamefont
  {F.}~\bibnamefont {Jelezko}},\ }\href@noop {} {\bibfield  {journal} {\bibinfo
   {journal} {Appl.\ Phys.\ Lett.}\ }\textbf {\bibinfo {volume} {103}},\
  \bibinfo {pages} {193118} (\bibinfo {year} {2013})}\BibitemShut {NoStop}%
\bibitem [{\citenamefont {Antonov}\ \emph {et~al.}(2014)\citenamefont
  {Antonov}, \citenamefont {H{\"a}u{\ss}ermann}, \citenamefont {Aird},
  \citenamefont {Roth}, \citenamefont {Trebin}, \citenamefont {M{\"u}ller},
  \citenamefont {McGuinness}, \citenamefont {Jelezko}, \citenamefont
  {Yamamoto}, \citenamefont {Isoya}, \citenamefont {Pezzagna}, \citenamefont
  {Meijer},\ and\ \citenamefont {Wrachtrup}}]{AHA+14}%
  \BibitemOpen
  \bibfield  {author} {\bibinfo {author} {\bibfnamefont {D.}~\bibnamefont
  {Antonov}}, \bibinfo {author} {\bibfnamefont {T.}~\bibnamefont
  {H{\"a}u{\ss}ermann}}, \bibinfo {author} {\bibfnamefont {A.}~\bibnamefont
  {Aird}}, \bibinfo {author} {\bibfnamefont {J.}~\bibnamefont {Roth}}, \bibinfo
  {author} {\bibfnamefont {H.-R.}\ \bibnamefont {Trebin}}, \bibinfo {author}
  {\bibfnamefont {C.}~\bibnamefont {M{\"u}ller}}, \bibinfo {author}
  {\bibfnamefont {L.}~\bibnamefont {McGuinness}}, \bibinfo {author}
  {\bibfnamefont {F.}~\bibnamefont {Jelezko}}, \bibinfo {author} {\bibfnamefont
  {T.}~\bibnamefont {Yamamoto}}, \bibinfo {author} {\bibfnamefont
  {J.}~\bibnamefont {Isoya}}, \bibinfo {author} {\bibfnamefont
  {S.}~\bibnamefont {Pezzagna}}, \bibinfo {author} {\bibfnamefont
  {J.}~\bibnamefont {Meijer}}, \ and\ \bibinfo {author} {\bibfnamefont
  {J.}~\bibnamefont {Wrachtrup}},\ }\href@noop {} {\bibfield  {journal}
  {\bibinfo  {journal} {Appl.\ Phys.\ Lett.}\ }\textbf {\bibinfo {volume}
  {104}},\ \bibinfo {pages} {012105} (\bibinfo {year} {2014})}\BibitemShut
  {NoStop}%
\bibitem [{\citenamefont {Sze}\ and\ \citenamefont {Lee}(2012)}]{SL12}%
  \BibitemOpen
  \bibfield  {author} {\bibinfo {author} {\bibfnamefont {S.~M.}\ \bibnamefont
  {Sze}}\ and\ \bibinfo {author} {\bibfnamefont {M.~K.}\ \bibnamefont {Lee}},\
  }\href@noop {} {\emph {\bibinfo {title} {Semiconductor {D}evices: {P}hysics
  and {T}echnology (3rd Ed.)}}}\ (\bibinfo  {publisher} {Wiley, Hoboken},\
  \bibinfo {year} {2012})\BibitemShut {NoStop}%
\bibitem [{\citenamefont {Cui}\ \emph {et~al.}(2015)\citenamefont {Cui},
  \citenamefont {Greenspon}, \citenamefont {Ohno}, \citenamefont {Myers},
  \citenamefont {Bleszynski~Jayich}, \citenamefont {Awschalom},\ and\
  \citenamefont {Hu}}]{CGO+15}%
  \BibitemOpen
  \bibfield  {author} {\bibinfo {author} {\bibfnamefont {S.}~\bibnamefont
  {Cui}}, \bibinfo {author} {\bibfnamefont {A.~S.}\ \bibnamefont {Greenspon}},
  \bibinfo {author} {\bibfnamefont {K.}~\bibnamefont {Ohno}}, \bibinfo {author}
  {\bibfnamefont {B.~A.}\ \bibnamefont {Myers}}, \bibinfo {author}
  {\bibfnamefont {A.~C.}\ \bibnamefont {Bleszynski~Jayich}}, \bibinfo {author}
  {\bibfnamefont {D.~D.}\ \bibnamefont {Awschalom}}, \ and\ \bibinfo {author}
  {\bibfnamefont {E.~L.}\ \bibnamefont {Hu}},\ }\href@noop {} {\bibfield
  {journal} {\bibinfo  {journal} {Nano Lett.}\ }\textbf {\bibinfo {volume}
  {15}},\ \bibinfo {pages} {2887} (\bibinfo {year} {2015})}\BibitemShut
  {NoStop}%
\bibitem [{\citenamefont {de~Oliveira}\ \emph {et~al.}(2015)\citenamefont
  {de~Oliveira}, \citenamefont {Momenzadeh}, \citenamefont {Wang},
  \citenamefont {Konuma}, \citenamefont {Markham}, \citenamefont {Edmonds},
  \citenamefont {Denisenko},\ and\ \citenamefont {Wrachtrup}}]{OMW+15}%
  \BibitemOpen
  \bibfield  {author} {\bibinfo {author} {\bibfnamefont {F.~F.}\ \bibnamefont
  {de~Oliveira}}, \bibinfo {author} {\bibfnamefont {S.~A.}\ \bibnamefont
  {Momenzadeh}}, \bibinfo {author} {\bibfnamefont {Y.}~\bibnamefont {Wang}},
  \bibinfo {author} {\bibfnamefont {M.}~\bibnamefont {Konuma}}, \bibinfo
  {author} {\bibfnamefont {M.}~\bibnamefont {Markham}}, \bibinfo {author}
  {\bibfnamefont {A.~M.}\ \bibnamefont {Edmonds}}, \bibinfo {author}
  {\bibfnamefont {A.}~\bibnamefont {Denisenko}}, \ and\ \bibinfo {author}
  {\bibfnamefont {J.}~\bibnamefont {Wrachtrup}},\ }\href@noop {} {\bibfield
  {journal} {\bibinfo  {journal} {Appl.\ Phys.\ Lett.}\ }\textbf {\bibinfo
  {volume} {107}},\ \bibinfo {pages} {073107} (\bibinfo {year}
  {2015})}\BibitemShut {NoStop}%
\bibitem [{\citenamefont {Zeigler}()}]{SRIM}%
  \BibitemOpen
  \bibfield  {author} {\bibinfo {author} {\bibfnamefont {J.~F.}\ \bibnamefont
  {Zeigler}},\ }\href@noop {} {\emph {\bibinfo {title} {The {S}topping and
  {R}ange of {I}ons in {M}atter, {SRIM}-2013}}},\ \bibinfo {note}
  {http://www.srim.org/}\BibitemShut {NoStop}%
\bibitem [{\citenamefont {Zeigler}(2004)}]{Z04}%
  \BibitemOpen
  \bibfield  {author} {\bibinfo {author} {\bibfnamefont {J.~F.}\ \bibnamefont
  {Zeigler}},\ }\href@noop {} {\bibfield  {journal} {\bibinfo  {journal}
  {Nucl.\ Instrum.\ Methods Phys.\ Res.\ B}\ }\textbf {\bibinfo {volume}
  {219-220}},\ \bibinfo {pages} {1027} (\bibinfo {year} {2004})}\BibitemShut
  {NoStop}%
\bibitem [{\citenamefont {Fu}\ \emph {et~al.}(2010)\citenamefont {Fu},
  \citenamefont {Santori}, \citenamefont {Barclay},\ and\ \citenamefont
  {Beausoleil}}]{FSB+10}%
  \BibitemOpen
  \bibfield  {author} {\bibinfo {author} {\bibfnamefont {K.-M.~C.}\
  \bibnamefont {Fu}}, \bibinfo {author} {\bibfnamefont {C.}~\bibnamefont
  {Santori}}, \bibinfo {author} {\bibfnamefont {P.~E.}\ \bibnamefont
  {Barclay}}, \ and\ \bibinfo {author} {\bibfnamefont {R.~G.}\ \bibnamefont
  {Beausoleil}},\ }\href@noop {} {\bibfield  {journal} {\bibinfo  {journal}
  {Appl.\ Phys.\ Lett.}\ }\textbf {\bibinfo {volume} {96}},\ \bibinfo {pages}
  {121907} (\bibinfo {year} {2010})}\BibitemShut {NoStop}%
\bibitem [{\citenamefont {Gali}, \citenamefont {Fyta},\ and\ \citenamefont
  {Kaxiras}(2008)}]{GFK08}%
  \BibitemOpen
  \bibfield  {author} {\bibinfo {author} {\bibfnamefont {A.}~\bibnamefont
  {Gali}}, \bibinfo {author} {\bibfnamefont {M.}~\bibnamefont {Fyta}}, \ and\
  \bibinfo {author} {\bibfnamefont {E.}~\bibnamefont {Kaxiras}},\ }\href@noop
  {} {\bibfield  {journal} {\bibinfo  {journal} {Phys.\ Rev.\ B}\ }\textbf
  {\bibinfo {volume} {77}},\ \bibinfo {pages} {155206} (\bibinfo {year}
  {2008})}\BibitemShut {NoStop}%
\bibitem [{\citenamefont {Mizuochi}\ \emph {et~al.}(2009)\citenamefont
  {Mizuochi}, \citenamefont {Neumann}, \citenamefont {Rempp}, \citenamefont
  {Beck}, \citenamefont {Jacques}, \citenamefont {Siyushev}, \citenamefont
  {Nakamura}, \citenamefont {Twitchen}, \citenamefont {Watanabe}, \citenamefont
  {Yamasaki}, \citenamefont {Jelezko},\ and\ \citenamefont
  {Wrachtrup}}]{MNR+09}%
  \BibitemOpen
  \bibfield  {author} {\bibinfo {author} {\bibfnamefont {N.}~\bibnamefont
  {Mizuochi}}, \bibinfo {author} {\bibfnamefont {P.}~\bibnamefont {Neumann}},
  \bibinfo {author} {\bibfnamefont {F.}~\bibnamefont {Rempp}}, \bibinfo
  {author} {\bibfnamefont {J.}~\bibnamefont {Beck}}, \bibinfo {author}
  {\bibfnamefont {V.}~\bibnamefont {Jacques}}, \bibinfo {author} {\bibfnamefont
  {P.}~\bibnamefont {Siyushev}}, \bibinfo {author} {\bibfnamefont
  {K.}~\bibnamefont {Nakamura}}, \bibinfo {author} {\bibfnamefont {D.~J.}\
  \bibnamefont {Twitchen}}, \bibinfo {author} {\bibfnamefont {H.}~\bibnamefont
  {Watanabe}}, \bibinfo {author} {\bibfnamefont {S.}~\bibnamefont {Yamasaki}},
  \bibinfo {author} {\bibfnamefont {F.}~\bibnamefont {Jelezko}}, \ and\
  \bibinfo {author} {\bibfnamefont {J.}~\bibnamefont {Wrachtrup}},\ }\href@noop
  {} {\bibfield  {journal} {\bibinfo  {journal} {Phys.\ Rev.\ B}\ }\textbf
  {\bibinfo {volume} {80}},\ \bibinfo {pages} {041201} (\bibinfo {year}
  {2009})}\BibitemShut {NoStop}%
\bibitem [{\citenamefont {Hahn}(1950)}]{H50}%
  \BibitemOpen
  \bibfield  {author} {\bibinfo {author} {\bibfnamefont {E.~L.}\ \bibnamefont
  {Hahn}},\ }\href@noop {} {\bibfield  {journal} {\bibinfo  {journal} {Phys.\
  Rev.}\ }\textbf {\bibinfo {volume} {80}},\ \bibinfo {pages} {580} (\bibinfo
  {year} {1950})}\BibitemShut {NoStop}%
\bibitem [{\citenamefont {Childress}\ \emph {et~al.}(2006)\citenamefont
  {Childress}, \citenamefont {Dutt}, \citenamefont {Taylor}, \citenamefont
  {Zibrov}, \citenamefont {Jelezko}, \citenamefont {Wrachtrup}, \citenamefont
  {Hemmer},\ and\ \citenamefont {Lukin}}]{CDT+06}%
  \BibitemOpen
  \bibfield  {author} {\bibinfo {author} {\bibfnamefont {L.}~\bibnamefont
  {Childress}}, \bibinfo {author} {\bibfnamefont {M.~V.~G.}\ \bibnamefont
  {Dutt}}, \bibinfo {author} {\bibfnamefont {J.~M.}\ \bibnamefont {Taylor}},
  \bibinfo {author} {\bibfnamefont {A.~S.}\ \bibnamefont {Zibrov}}, \bibinfo
  {author} {\bibfnamefont {F.}~\bibnamefont {Jelezko}}, \bibinfo {author}
  {\bibfnamefont {J.}~\bibnamefont {Wrachtrup}}, \bibinfo {author}
  {\bibfnamefont {P.~R.}\ \bibnamefont {Hemmer}}, \ and\ \bibinfo {author}
  {\bibfnamefont {M.~D.}\ \bibnamefont {Lukin}},\ }\href@noop {} {\bibfield
  {journal} {\bibinfo  {journal} {Science}\ }\textbf {\bibinfo {volume}
  {314}},\ \bibinfo {pages} {281} (\bibinfo {year} {2006})}\BibitemShut
  {NoStop}%
\bibitem [{\citenamefont {Pham}\ \emph {et~al.}(2016)\citenamefont {Pham},
  \citenamefont {DeVience}, \citenamefont {Casola}, \citenamefont {Lovchinsky},
  \citenamefont {Sushkov}, \citenamefont {Bersin}, \citenamefont {Lee},
  \citenamefont {Urbach}, \citenamefont {Cappellaro}, \citenamefont {Park},
  \citenamefont {Yacoby}, \citenamefont {Lukin},\ and\ \citenamefont
  {Walsworth}}]{PDC+16}%
  \BibitemOpen
  \bibfield  {author} {\bibinfo {author} {\bibfnamefont {L.~M.}\ \bibnamefont
  {Pham}}, \bibinfo {author} {\bibfnamefont {S.~J.}\ \bibnamefont {DeVience}},
  \bibinfo {author} {\bibfnamefont {F.}~\bibnamefont {Casola}}, \bibinfo
  {author} {\bibfnamefont {I.}~\bibnamefont {Lovchinsky}}, \bibinfo {author}
  {\bibfnamefont {A.~O.}\ \bibnamefont {Sushkov}}, \bibinfo {author}
  {\bibfnamefont {E.}~\bibnamefont {Bersin}}, \bibinfo {author} {\bibfnamefont
  {J.}~\bibnamefont {Lee}}, \bibinfo {author} {\bibfnamefont {E.}~\bibnamefont
  {Urbach}}, \bibinfo {author} {\bibfnamefont {P.}~\bibnamefont {Cappellaro}},
  \bibinfo {author} {\bibfnamefont {H.}~\bibnamefont {Park}}, \bibinfo {author}
  {\bibfnamefont {A.}~\bibnamefont {Yacoby}}, \bibinfo {author} {\bibfnamefont
  {M.}~\bibnamefont {Lukin}}, \ and\ \bibinfo {author} {\bibfnamefont {R.~L.}\
  \bibnamefont {Walsworth}},\ }\href@noop {} {\bibfield  {journal} {\bibinfo
  {journal} {Phys.\ Rev.\ B}\ }\textbf {\bibinfo {volume} {93}},\ \bibinfo
  {pages} {045425} (\bibinfo {year} {2016})}\BibitemShut {NoStop}%
\bibitem [{\citenamefont {Loretz}\ \emph {et~al.}(2016)\citenamefont {Loretz},
  \citenamefont {Boss}, \citenamefont {Rosskopf}, \citenamefont {Mamin},
  \citenamefont {Rugar},\ and\ \citenamefont {Degen}}]{LBR+16}%
  \BibitemOpen
  \bibfield  {author} {\bibinfo {author} {\bibfnamefont {M.}~\bibnamefont
  {Loretz}}, \bibinfo {author} {\bibfnamefont {J.~M.}\ \bibnamefont {Boss}},
  \bibinfo {author} {\bibfnamefont {T.}~\bibnamefont {Rosskopf}}, \bibinfo
  {author} {\bibfnamefont {H.~J.}\ \bibnamefont {Mamin}}, \bibinfo {author}
  {\bibfnamefont {D.}~\bibnamefont {Rugar}}, \ and\ \bibinfo {author}
  {\bibfnamefont {C.~L.}\ \bibnamefont {Degen}},\ }\href@noop {} {\bibfield
  {journal} {\bibinfo  {journal} {Phys.\ Rev.\ X}\ }\textbf {\bibinfo {volume}
  {5}},\ \bibinfo {pages} {021009} (\bibinfo {year} {2016})}\BibitemShut
  {NoStop}%
\bibitem [{\citenamefont {de~Oliveira}\ \emph {et~al.}()\citenamefont
  {de~Oliveira}, \citenamefont {Antonov}, \citenamefont {Wang}, \citenamefont
  {Neumann}, \citenamefont {Momenzadeh}, \citenamefont {H{\"a}u{\ss}ermann},
  \citenamefont {Pasquarelli}, \citenamefont {Denisenko},\ and\ \citenamefont
  {Wrachtrup}}]{dOAW+17}%
  \BibitemOpen
  \bibfield  {author} {\bibinfo {author} {\bibfnamefont {F.~F.}\ \bibnamefont
  {de~Oliveira}}, \bibinfo {author} {\bibfnamefont {D.}~\bibnamefont
  {Antonov}}, \bibinfo {author} {\bibfnamefont {Y.}~\bibnamefont {Wang}},
  \bibinfo {author} {\bibfnamefont {P.}~\bibnamefont {Neumann}}, \bibinfo
  {author} {\bibfnamefont {S.~A.}\ \bibnamefont {Momenzadeh}}, \bibinfo
  {author} {\bibfnamefont {T.}~\bibnamefont {H{\"a}u{\ss}ermann}}, \bibinfo
  {author} {\bibfnamefont {A.}~\bibnamefont {Pasquarelli}}, \bibinfo {author}
  {\bibfnamefont {A.}~\bibnamefont {Denisenko}}, \ and\ \bibinfo {author}
  {\bibfnamefont {J.}~\bibnamefont {Wrachtrup}},\ }\href@noop {} {}\bibinfo
  {note} {{a}rXiv:1701.07055v1 (unpublished)}\BibitemShut {NoStop}%
\end{thebibliography}%
\end{document}